\documentclass[aps,pra,twocolumn,groupedaddress,showpacs,showkeys]{revtex4}

\usepackage{graphicx}
\begin{document}

\newcommand{\be}{\begin{equation}}
\newcommand{\ee}{\end{equation}}

\title{A generalized Kramers-Kronig transform for Casimir effect computations}

\author{ Giuseppe Bimonte}
\email[Bimonte@na.infn.it]
\affiliation{Dipartimento di Scienze Fisiche Universit\`{a} di
Napoli Federico II Complesso Universitario MSA, Via Cintia
I-80126 Napoli Italy and INFN Sezione di Napoli, ITALY\\
}

\date{\today}

\begin{abstract}
Recent  advances in experimental techniques  now permit to measure
the Casimir force with unprecedented precision. In order to
achieve a comparable precision in the theoretical prediction of
the force, it is necessary to accurately determine the electric
permittivity of the materials constituting the plates  along the
imaginary frequency axis. The latter quantity is not directly
accessible to experiments, but it can be determined via dispersion
relations from experimental optical data. In the experimentally
important case of conductors, however, a serious drawback of the
standard dispersion relations commonly used for this purpose, is
their strong dependence on the chosen low-frequency extrapolation
of the experimental optical data, which introduces a significant
and not easily controllable  uncertainty in the result. In this
paper we show that a simple modification of the standard
dispersion relations, involving suitable analytic window
functions, resolves this difficulty, making it possible to
reliably determine the electric permittivity at imaginary
frequencies solely using experimental optical data in the
frequency interval where they are available, without any need of
uncontrolled   data extrapolations.
\end{abstract}


\maketitle

\section{INTRODUCTION}

One of the most intriguing predictions of Quantum electrodynamics
is the existence of irreducible vacuum fluctuations of the
electromagnetic (e.m.) field. It was Casimir's fundamental
discovery \cite{casimir} to realize that   this purely quantum
phenomenon was not confined to the atomic scale, as in the Lamb
shift, but would rather manifest itself also at the macroscopic
scale, in the form of a force of attraction between two discharged
plates. For the idealized case of two perfectly reflecting
plane-parallel plates at zero temperature, placed at a distance
$a$ in vacuum, Casimir obtained the following remarkably simple
estimate of the force per unit area \be F_{C}=\frac{\pi^2 \hbar c
}{240 \;a^4}\;.\label{paral}\ee An important step forward was made
a few years later by Lifshitz and co-workers \cite{lifs}, who
obtained a formula for the force between two homogeneous
dielectric plane-parallel slabs, at finite temperature. In this
theory, of macroscopic character, the material properties of the
slabs were fully characterized in terms of the respective
frequency dependent electric permittivities $\epsilon(\omega)$,
accounting for the dispersive and dissipative properties of real
materials. In this way, it was  possible for the first time to
investigate the influence of material properties on the magnitude
of the Casimir force.

Over ten years ago, a series of brilliant experiments
\cite{lamor,roy} exploiting modern experimental techniques
provided the definitive demonstration of the Casimir effect. These
now historical experiments spurred enormous interest in the
Casimir effect, and were soon followed by many other experiments.
The subsequent experiments were aimed at diverse objectives. Some
of them explored new geometries: while the works \cite{lamor,roy}
used a sphere-plate setup, the original planar geometry
investigated by  Casimir was adopted in the experiment
\cite{Bressi}, and a setup with crossed cylinders was  considered
in \cite{ederth}. The important issue of the non trivial geometry
dependence of the Casimir effect is also being pursued
experimentally, using elaborate micro-patterned surfaces
\cite{Bao}. Other experiments aimed at demonstrating new possible
uses of the Casimir force, like for example the actuation of
micromachines \cite{capasso}, or at demonstrating the possibility
of a large modulation of the Casimir force \cite{chen,iannuzzi},
which could also result in interesting technological applications.
There are also experiments using superconducting Casimir cavities,
that aim at measuring the change of the Casimir energy across the
superconducting phase transition \cite{bimonte}. The experiments
performed in the last ten years are just too numerous to mention
them all here. For an updated account we refer the reader to the
very recent review paper \cite{Mohid}.

Apart from exploring new manifestations of the Casimir effect, a
large experimental effort is presently being made also to increase
the precision of Casimir force measurements, in simple geometries.
Already in the early experiment \cite{roy} a precision upto one
percent was obtained. More recently, a series of experiments with
microtorsional oscillators \cite{decca} reached an amazing
precision of 0.2 percent. The reader may wonder what is the
interest in achieving such a high precision in this kind of
experiments. There are several reasons why this is important. On
one hand, in the theory of dispersion forces  puzzling conceptual
problems have recently emerged that are connected with the
contribution of free charges to the thermal Casimir force, whose
resolution crucially depends on the precision of the
theory-experiment comparison  \cite{Mohid}. On the other hand, the
ability to accurately  determine the Casimir force is also
important for the purpose of obtaining stronger constraints on
hypothetical long-range forces predicted by certain theoretical
scenarios going beyond the Standard Model of particle physics
\cite{Mohid}.

The remarkable   precision achieved in the most recent experiments
poses a challenging demand on the theorist: is it possible to
predict the magnitude of the Casimir force with a comparable level
of precision, say of one percent? Assessing the theoretical error
affecting present estimates of the Casimir force is a difficult
problem indeed, because many different factors must be taken into
account \cite{Mohid}. Consider the typical experimental setting of
most of the current experiments, where   the Casimir force is
measured between two bodies covered with gold, placed in vacuum at
a distance of a (few) hundred nanometers. In this separation
range, the main factor to consider is the finite penetration depth
of electromagnetic fields into the gold layer \footnote{In typical
experiments, the thickness of the gold layers covering the bodies
is large enough, that one can neglect the material of the
substrate, and treat the bodies as if they were made just of
gold.}, resulting from the finite conductivity of gold. The tool
to analyze the influence of such material properties as the
conductivity on the Casimir effect is provided by Lifshitz theory
\cite{lifs}. This theory shows that for a separation of 100 nm,
the finite conductivity of gold determines a reduction in the
magnitude of the Casimir force of about fifty percent in
comparison with the perfect metal result \cite{lambr}. Much
smaller corrections, that must nevertheless be considered if the
force is to be estimated with percent precision, arise from the
finite temperature of the plates and from their surface roughness.
Moreover, geometric effects resulting from the actual shape of the
plates should be considered. We should also mention that the
magnitude of residual electrostatic forces between the plates,
resulting from contact potentials and patch effects, must be
carefully accounted for. For a discussion of all these issues,
which received much attention in the recent literature on the
Casimir effect, we again address the reader to Ref. \cite{Mohid}.
See also the recent work \cite{rudy}.

In this paper, we focus our attention on the influence of the
optical properties of the plates which, as explained above,  is by
far the most relevant factor to consider.
As we pointed out earlier, in Lifshitz theory the optical
properties of the plates enter via the frequency-dependent
electric permittivity $\epsilon(\omega)$ of the material
constituting plates. In order to obtain an accurate prediction of
the force, it is therefore of the utmost importance to use
accurate data for the electric permittivity.   The common practice
adopted in all recent Casimir experiments with gold surfaces is to
use tabulated data for gold (most of the times those quoted in
Refs. \cite{palik}), suitably extrapolated at low frequencies,
where optical data are not available, by simple analytic models
(like the Drude model or the so-called generalized plasma model).
However, already ten years ago Lamoreaux observed \cite{lamor2}
that using tabulated data to obtain an accurate prediction of the
Casimir force
 may not be a  reliable practice, since
optical properties of gold films may vary significantly from
sample to sample, depending on the conditions of deposition. The
same author stressed the importance of measuring the optical data
of the films actually used in the force measurements, in the
frequency range that is relevant for the Casimir force. The
importance of this point was further stressed in \cite{Piro} and
received clear experimental support in a recent paper
\cite{sveto}, where the optical properties of several gold films
of different thicknesses, and prepared by different procedures,
were measured ellipsometrically in a wide range of wavelengths,
from 0.14 to 33 microns, and it was found that the frequency
dependent electric permittivity changes significantly from sample
to sample. By using the zero-temperature Lifshitz formula, the
authors estimated that the observed sample dependence of the
electric permittivity implies a variation in the theoretical value
of the Casimir force, from one sample to another, easily as large
as ten percent, for separations around 100 nm. It was concluded
that in order to achieve a theoretical accuracy  better than ten
percent in the prediction of the Casimir force, it is necessary to
determine the optical properties of the films actually used in the
experiment of interest.

The aim of this paper is to improve the mathematical procedure
that is actually needed to obtain reliable estimates of the
Casimir force, starting from experimental optical data on the
material of the plates, like those presented in Ref. \cite{sveto}.
The necessity of such an improvement stems from the very simple
and unavoidable fact that experimental optical data are never
available in the entire frequency domain, but are always
restricted to a finite frequency interval $\omega_{\rm min} <
\omega < \omega_{\rm max}$.
To see why this constitutes problem we recall that Lifshitz
formula, routinely used to interpret current experiments,
expresses the Casimir force between two parallel plates as an
integral over {\it imaginary} frequencies ${\rm i} \xi$ of a
quantity involving the dielectric permittivities of the plates
$\epsilon({\rm i \xi})$. For finite temperature, the continuous
frequency integration is replaced by a sum over discrete so-called
Matsubara frequencies $\omega_n={\rm i} \xi_n$, where $\xi_n=2 \pi
n k_B T /\hbar$, with $n$ a non-negative integer, and $T$ the
temperature of the plates. In any case, whatever the temperature,
one needs to evaluate the permittivity of the plates for certain
imaginary frequencies. We note that, in principle, recourse to
imaginary frequencies is  not mandatory because it is possible to
rewrite Lifshitz formula in a mathematically equivalent form,
involving an integral over the real frequency axis. In this case
however the integrand becomes a rapidly oscillating function of
the frequency, which hampers any possibility of numerical
evaluation. In practice, the real-frequency form of Lifshitz
formula is never used, and only its imaginary-frequency version is
considered. We remark that   occurrence of imaginary frequencies
in the expression of the Casimir force, is a general feature of
all recent formalisms, extending Lifshitz theory to non-planar
geometries \cite{emig,kenneth,lambr2}. The problem is that the
electric permittivity $\epsilon(i \xi)$ at imaginary frequencies
cannot be measured directly by any experiment. The only way to
determine  it by means of dispersion relations, which allow to
express $\epsilon(i \xi)$ in terms of the observable
real-frequency electric permittivity $\epsilon(\omega)$. In the
standard version of dispersion relations \cite{lifs}, adopted so
far in all works on the Casimir effect, $\epsilon({\rm i} \xi)-1$
is expressed in terms of an integral  of a quantity involving the
imaginary part $\epsilon''(\omega)$ of the electric permittivity:
\be \epsilon({\rm i} \xi)-1=\frac{2}{\pi}\int_0^{\infty} d \omega
\frac{\omega\,
\epsilon''(\omega)}{\omega^2+\xi^2}\;.\label{disp}\ee The above
formula shows that, in principle, a determination of
$\epsilon({\rm i} \xi)$ requires knowledge of $\epsilon''(\omega)$
at all frequencies while, as we said earlier, optical data are
available only in some interval $\omega_{\rm min} < \omega <
\omega_{\rm max}$. In practice, the problem is not so serious on
the high frequency side, because the fall-off properties of
$\epsilon''(\omega)$ at high frequencies, together with the
$\omega^2$ factor in the denominator of the integrand, ensure that
the error made by truncating the integral at a suitably large
frequency $\omega_{\rm max}$ is small, provided that $\omega_{\rm
max}$ is large enough. Typically, an $\omega_{\rm max}$  larger
than, say, $15 c/(2 a)$, is good enough for practical purposes.
Things are not so easy though on the low frequency side. In the
case of insulators, optical data are typically available until
frequencies $\omega_{\min}$ much smaller than the frequencies of
all resonances of the medium. Because of this,
$\epsilon''(\omega)$ is almost zero for $\omega < \omega_{\rm
min}$, and therefore the error made by truncating the integral at
$\omega_{\rm min}$ is again negligible. Problems arise however in
the case of ohmic conductors, because then $\epsilon''(\omega)$
has a $1/\omega$ singularity at $\omega=0$. As a result
$\epsilon''(\omega)$ becomes extremely large at low frequencies,
in such a way that the integral in Eq. (\ref{disp}) receives a
very large contribution from low frequencies. For typical values
of $\omega_{\rm min}$ that can be reached in practice (for example
for gold, the tabulated data in \cite{palik} begin at $\omega_{\rm
min }=125$ meV$/\hbar$, while the data of \cite{sveto} start at 38
meV$/\hbar$) truncation of the integral at $\omega_{\rm min}$
results in a large error. The traditional remedy to this problem
is to make some analytical extrapolation of the data, typically
based on Drude model fits of the low-frequency region of data,
from $\omega_{\rm min}$ to zero, and then use the extrapolation to
estimate the contribution of the integral in the interval $0 <
\omega <\omega_{\rm min}$ where data are not directly available.
It is important to observe that this contribution is usually very
large. For example, even in the case of Ref. \cite{sveto},
the relative contribution of
the extrapolation is about fifty percent of the total value of the
integral, in the entire range of imaginary frequencies that are
needed for estimating the Casimir force.

Clearly, this procedure is not very satisfying. The use of
analytical extrapolations of the data introduces an uncertainty in
the obtained values of $\epsilon({\rm i} \xi)$, that is not easy
to quantify. The result may in fact depend a lot on the form of
the extrapolation, and there is no guarantee that the chosen
expression is good enough. Consider for example Ref. \cite{sveto},
which   constitutes the most accurate existing work on this
problem. It was found there that the simple Drude model does not
fit so well the data of all samples, making it necessary to
improve it by the inclusion  of an additional Lorentz oscillator.
Moreover, it was found that for each sample the Drude parameters
extracted from the data depended on the used fitting procedure,
and were inconsistent which each other within the estimated
errors, which is again an indication of the probable inadequacy of
the analytical expression chosen for the interpolation. This state
of things led us to investigate if it possible to determine
accurately $\epsilon({\rm i \xi})$ solely on the basis of
available optical data, without making recourse to data
extrapolations. We shall see below that this is indeed possible,
provided that Eq. (\ref{disp}) is suitably modified, in a way that
involves multiplying the integrand by an appropriate analytical
window function $f(\omega)$, which suppresses the contribution of
frequencies not belonging to the  interval $\omega_{\rm min} <
\omega < \omega_{\rm max}$. As a result of this modification, the
error made by truncating the integral to the frequency range
$\omega_{\rm min} < \omega < \omega_{\rm max}$ can be made
negligible at both ends of the integration domain, rendering
unnecessary any extrapolation of the optical data outside the
interval where they are available.
The procedure outlined in this paper should allow to better
evaluate the theoretical uncertainty of Casimir force estimates
resulting from experimental errors in the optical data.

The plan of the paper is as follows: in Sec. II we derive a
generalized dispersion relation for $\epsilon({\rm i} \xi)$,
involving analytic window functions $f(z)$, and we provide a
simple choice for the window functions. In Sec III we present the
results of a numerical simulation of our window functions, for the
experimentally relevant case of gold, and in Sec IV we estimate
numerically the error on the Casimir pressure resulting from the
use of our window functions. Sec V contains our conclusions and a
discussion of the results.

\section{Generalized dispersion relations with window-functions }

As it it well known \cite{lifs},  analyticity properties satisfied
by the electric permittivity $\epsilon(\omega)$ of any causal
medium (and more in general by any causal response function, the
magnetic permeability $\mu(\omega)$ being another example) imply
certain integral relations between the real part
$\epsilon'(\omega)$ and imaginary part $\epsilon''(\omega)$ of
$\epsilon(\omega)$, known as Kramers-Kronig or dispersion
relations. The dispersion relation of interest to us is the one
that permits to express the value $\epsilon({\rm i} \xi)$ of the
response function at some imaginary frequency ${\rm i} \xi$ in
terms of an integral along the positive frequency axis, involving
$\epsilon''(\omega)$. It is convenient to briefly review here the
simple derivation of this important result, which is an easy
exercise in contour integration. For our purposes, it is more
convenient to start from an arbitrary complex function $u(z)$,
with the following properties: $u(z)$ is analytic in the upper
complex plane ${\cal C}^+=\{z: {\rm Im}(z)>0\}$, fall's off to
zero for large $|z|$ like some power of $|z|$, and admits at most
a simple pole at $\omega=0$.  Consider now the closed integration
contour $\Gamma$ obtained  by closing in the upper complex plane
the positively oriented real axis, and let $z_0$ be any complex
number in ${\cal C}^+$. It is then a simple matter to verify the
identity: \be \int_{\Gamma} dz \frac{z \, u(z)}{z^2-z_0^2}= {\rm
i} \pi u(z_0)\;.\label{intuno}\ee   The assumed fall-off property
of $u(z)$ ensures that the half-circle of infinite radius forming
$\Gamma$ contributes nothing to the integral, and then from Eq.
(\ref{intuno}) we find: \be
  u(z_0)=\frac{1}{{\rm i} \pi}\int_{-\infty}^{\infty} d\omega \frac{\omega
\, u(\omega)}{\omega^2-z_0^2}\;.\label{intdue}\ee Consider now a
purely imaginary complex number $z_0={\rm i} \xi$, and assume in
addition that  along the real axis $u(\omega)$ satisfies the
symmetry property $u(-\omega)=u^*(\omega)$. From Eq.
(\ref{intdue}) we then find: \be u({\rm i} \xi)=\frac{2}{
\pi}\int_0^{\infty} d\omega \frac{\omega \,
u''(\omega)}{\omega^2+\xi^2}\;,\label{gendisp}\ee which is the
desired result.

The standard dispersion relation Eq. (\ref{disp}) used to compute
the electric permittivity for imaginary frequencies is a special
case of the above relation, corresponding to choosing
$u(z)=\epsilon(z)-1$. We note that Eq. (\ref{disp}) is valid both
for insulators, which have a finite permittivity at zero
frequency, as well as for ohmic conductors, whose permittivity has
a $1/\omega$ singularity in the origin. As we explained in the
introduction  Eq. (\ref{disp}), even though perfectly correct from
a  mathematical standpoint,  has serious drawbacks, when it is
used to numerically estimate $\epsilon(i \xi)$ for ohmic
conductors, starting from optical data available only in some
interval $\omega_{\rm min} < \omega < \omega_{\rm max}$, because
the integral on the r.h.s. of Eq. (\ref{disp}) receives a large
contribution from frequencies near zero, where data are not
available. This difficulty   can however be overcome in a very
simple way, as we now explain. Consider a window function $f(z)$,
enjoying the following properties: $f(z)$ is analytic in ${\cal
C}^+$, it has no poles in ${\cal C}^+$ except possibly  a simple
pole at infinity, and satisfies the symmetry property \be
f(-z^*)=f^*(z)\;.\label{sym} \ee Consider now Eq. (\ref{gendisp}),
for $u(z)=f(z)(\epsilon(z)-1)$. Since  for any medium
$(\epsilon(z)-1)$ falls off like $z^{-2}$ at infinity
\cite{lifs2}, the quantity $u(z)$ falls off at least like $z^{-1}$
at infinity, and it satisfies all the properties required for Eq.
(\ref{gendisp}) to hold. For any $\xi$ such that $f(i \xi) \neq
0$, we then obtain the following generalized dispersion relation:
\be \epsilon({\rm i} \xi)-1= \frac{2}{\pi \, f({\rm i} \xi)}
\int_0^{\infty} d\omega \frac{\omega \, }{\omega^2+\xi^2}{\rm
Im}[f(\omega)(\epsilon(\omega)-1)]\;.\label{generdisp}\ee
We note that the above relation constitutes an exact result,
generalizing the standard dispersion relation Eq. (\ref{disp}), to
which it reduces with the choice $f(z)=1$. Another form of
dispersion relation, frequently used
 in the case of conductors or superconductors
\cite{bimonte,bimonte2} is obtained by taking $f(z)={\rm i}\,z$
into Eq. (\ref{generdisp}). Recalling the relation \cite{lifs2}
\be \epsilon(\omega)=1+\frac{4 \pi i}{\omega}
\,\sigma(\omega)\;,\ee it reads: \be \epsilon({\rm i} \xi)-1=
\frac{8}{  \xi} \int_0^{\infty} d\omega \frac{\omega \,
}{\omega^2+\xi^2}{\rm Im}\,[\sigma(\omega)]\;.\label{sigma}\ee The
above form is especially convenient in the case of
superconductors, because it avoids the $\delta(\omega)$
singularity characterizing the real part of the conductivity of
these materials \cite{bimonte2}.

We observe now, and this is the key point, that there is no reason
to restrict the choice of the function $f(z)$ to these two
possibilities. Indeed, we can take advantage of the freedom in the
choice of $f(z)$, to suppress the unwanted contribution of low
frequencies (as well as of high frequencies), where experimental
data on $\epsilon(\omega)$ are not available. In order to do that,
it is sufficient to choose a window function that goes to zero
fast enough for $\omega \rightarrow 0$, as well as for $\omega
\rightarrow \infty$. A convenient family of window functions which
do the job is the following: \be f(z)=A\, z^{2 p+1}\left[
\frac{1}{(z-w)^{2 q+1}} +\frac{1}{(z+w^*)^{2 q+1}}
\right]\;,\label{winfun}\ee where $w$ is an arbitrary complex
number such that ${\rm Im}(w) <0$, and $p$ and $q$ are integers
such that $p < q$. The constant $A$ is an irrelevant arbitrary
normalization constant, that drops out from the generalized
dispersion formula Eq. (\ref{generdisp}). As we see,  in the limit
$z \rightarrow 0$, these functions vanish like $z^{2p+1}$, and
therefore by taking sufficiently large values for $p$ we can
obtain suppression of low frequencies to any desired level. On the
other hand, for $z \rightarrow \infty$, $f(z)$ vanishes like
$z^{2(p-q)}$, and therefore by taking sufficiently large values of
$q$, we can obtain suppression of high frequencies.  Moreover, by
suitably choosing the free parameter $w$, we can also adjust the
range of frequencies that effectively contribute to the integral
on the r.h.s. of Eq. (\ref{generdisp}). In Figs. 1 and 2 we plot
the real and imaginary parts (in arbitrary units) of our window
functions $f(\omega)$, versus the frequency $\omega$ (expressed in
eV). The two curves displayed correspond to the choices
$p=1,\,q=2$ (dashed line) and $p=1,\,q=3$ (solid line). In both
cases, the parameter $w$ has the value $w=(1-2\, {\rm i}) \,{\rm
eV}/\hbar$.
\begin{figure}
\includegraphics{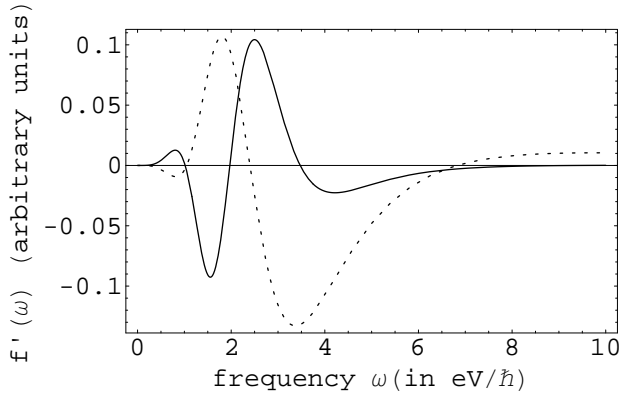}
\caption{\label{fig1}  Real part  $f'(\omega)$ (in arbitrary
units) of the window functions in Eq. (\ref{winfun}) versus
frequency $\omega$ (in eV/$\hbar$). The window parameters are
$p=1,\,q=2$ (dashed line), $p=1,\,q=3$ (solid line). In both
cases, the parameter $w$ has the value $w=(1-2\, {\rm i}) \,{\rm
eV}/\hbar$.}
\end{figure}
\begin{figure}
\includegraphics{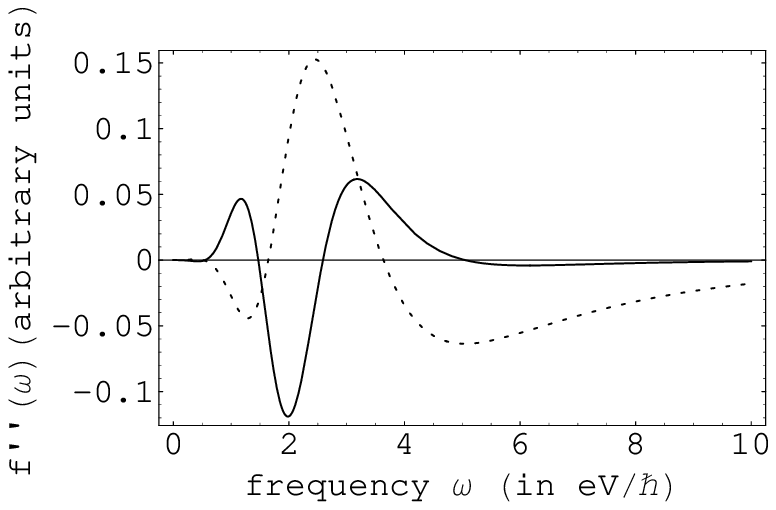}
\caption{\label{fig2}  Imaginary part $f''(\omega)$ (in arbitrary
units) of the window functions in Eq. (\ref{winfun}) versus
frequency $\omega$ (in eV/$\hbar$). The window parameters are
$p=1,\,q=2$ (dashed line), $p=1,\,q=3$ (solid line). In both
cases, the parameter $w$ has the value $w=(1-2\, {\rm i}) \,{\rm
eV}/\hbar$.}
\end{figure}
\begin{figure}
\includegraphics{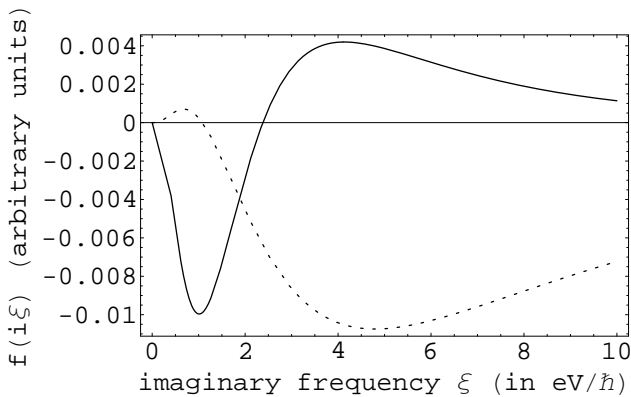}
\caption{\label{fig3} Plots (in arbitrary units) of the window
functions  $f({\rm i} \xi)$ in Eq. (\ref{winfun}) versus the
imaginary frequency $\xi$ (in eV/$\hbar$). The window parameters
are $p=1,\,q=2$ (dashed line), $p=1,\,q=3$ (solid line). In both
cases, the parameter $w$ has the $w=(1-2\, {\rm i}) \,{\rm
eV}/\hbar$.}
\end{figure}
We observe that along the real frequency axis, our window
functions have non-vanishing real and imaginary parts. This is not
a feature of our particular choice of the window functions,  but
it is an unavoidable consequence of our demand of analyticity on
$f(z)$. Indeed, for real frequencies $\omega$ the real and
imaginary parts of $f(\omega)$ are related to each other by the
usual Kramers-Kronig relations \cite{lifs} that hold for the
boundary values of analytic functions. In the case when $f(z)$
vanishes at infinity, they read: \be f'(\omega)=\frac{1}{\pi}{\rm
P} \int_{-\infty}^{\infty} d \xi\,
\frac{f''(\xi)}{\xi-\omega}\;,\label{reKK}\ee \be
f''(\omega)=-\frac{1}{\pi}{\rm P} \int_{-\infty}^{\infty} d \xi\,
\frac{f'(\xi)}{\xi-\omega}\;,\label{imKK}\ee where the symbol
${\rm P}$ in front of the integrals denotes the principal value.
These relation show that vanishing of $f'(\omega)$ implies that of
$f''(\omega)$ and viceversa, and therefore neither $f'(\omega)$
nor $f''(\omega)$ can be identically zero. By virtue of this
property of the window functions, it follows from Eq.
(\ref{generdisp}) that  both the real and imaginary parts of
$\epsilon(\omega)$ are needed  to evaluate $\epsilon({\rm i} \xi)$
(unless the standard choices $f(z) \equiv 1$ or $f(z)={\rm i}\,z$
are made). We also note (see Fig. 1 and 2) that the real and
imaginary parts of $f(\omega)$ do not have a definite sign. This
feature also is a general consequence of our key demand that
$f(z)$ vanishes in the origin, as it can be seen by taking
$\omega=0$ in Eqs. (\ref{reKK}) and (\ref{imKK}). Since the l.h.s.
of both equations are required to vanish, the integrand on the
r.h.s. cannot have a definite sign. Finally, in Fig 3 we show
plots of two of our window functions $f({\rm i} \xi)$, versus the
imaginary frequency $\xi$, expressed in eV, for the same two
choices of parameters of Fig. 1 and 2. It is important to observe
that the window functions $f(z)$ are real along the imaginary axis
(as it must be, as a consequence of the symmetry property Eq.
(\ref{sym})). However, the sign of $f({\rm i} \xi)$ is not
definite, and as a result of this  $f({\rm i} \xi)$ admits zeros
along the imaginary axis. When using Eq. (\ref{generdisp}) for
estimating $\epsilon({\rm i \xi})$ it is then important to choose
the window function such that none of its zeroes coincides with
the value of $\xi$ for which $\epsilon({\rm i \xi})$  is being
estimated.

\section{A numerical simulation}

In this Section, we perform a simple simulation to test the degree
of accuracy with which the quantity $\epsilon({\rm i} \xi)$ can be
reconstructed using our window functions, starting from data on
$\epsilon(\omega)$ referring to a finite frequency interval.  To
do that we can proceed as follows.

According to the standard dispersion relation  Eq. (\ref{disp}),
the quantity $\epsilon({\rm i} \xi)-1$ is equal to the integral on
the r.h.s. of Eq. (\ref{disp}). Following Refs. \cite{Piro,sveto},
we can split this integral into three pieces, as follows: \be
\frac{2}{\pi}\int_0^{\infty} d \omega \frac{\omega\,
\epsilon''(\omega)}{\omega^2+\xi^2}=I_{\rm low}(\xi)+I_{\rm
exp}(\xi)+I_{\rm high}(\xi)\;,\label{split}\ee where we set: \be
I_{\rm low}(\xi)=\frac{2}{\pi}\int_0^{\omega_{\rm min}} d \omega
\frac{\omega\, \epsilon''(\omega)}{\omega^2+\xi^2}\;,\ee \be
I_{\rm exp}(\xi)=\frac{2}{\pi}\int_{\omega_{\rm min}}^{\omega_{\rm
max}} d \omega \frac{\omega\,
\epsilon''(\omega)}{\omega^2+\xi^2}\;,\ee and \be I_{\rm
high}(\xi)=\frac{2}{\pi}\int_{\omega_{\rm max}}^{\infty} d \omega
\frac{\omega\, \epsilon''(\omega)}{\omega^2+\xi^2}\;.\ee By
construction, we obviously have: \be \epsilon({\rm i}
\xi)-1=I_{\rm low}(\xi)+I_{\rm exp}(\xi)+I_{\rm high}(\xi)\;.\ee
An analogous split can be performed in the integral on the r.h.s.
of the other standard dispersion relation involving the
conductivity Eq. (\ref{sigma}): \be  \frac{8}{  \xi}
\int_0^{\infty} d\omega \frac{\omega \, }{\omega^2+\xi^2}{\rm
Im}\,[\sigma(\omega)]=K_{\rm low}(\xi)+K_{\rm exp}(\xi)+K_{\rm
high}(\xi)\;,\label{splitsigma}\ee with an obvious meaning of the
symbols. Again, we have the identity: \be \epsilon({\rm i}
\xi)-1=K_{\rm low}(\xi)+K_{\rm exp}(\xi)+K_{\rm high}(\xi)\;.\ee
On the other hand, according to our generalized dispersion
relation Eq. (\ref{generdisp}), the quantity $\epsilon({\rm i}
\xi)-1$ is also equal to the integral on the r.h.s. of Eq.
(\ref{generdisp}). We can split this integral too in a way
analogous to Eq. (\ref{split}):
$$ \frac{2}{\pi \, f({\rm i} \xi)} \int_0^{\infty} d\omega
\frac{\omega \, }{\omega^2+\xi^2}{\rm
Im}[f(\omega)(\epsilon(\omega)-1)]$$ \be =J_{\rm
low}^{(p,q)}(\xi)+J_{\rm exp}^{(p,q)}(\xi)+J_{\rm
high}^{(p,q)}(\xi)\;,\ee where we set: \be J_{\rm
low}^{(p,q)}(\xi)=\frac{2}{\pi\, f({\rm i}
\xi)}\int_0^{\omega_{\rm min}} d\omega \frac{\omega \,
}{\omega^2+\xi^2}{\rm Im}[f(\omega)(\epsilon(\omega)-1)]\;,\ee \be
J_{\rm exp}^{(p,q)}(\xi)=\frac{2}{\pi\, f({\rm i}
\xi)}\int_{\omega_{\rm min}}^{\omega_{\rm max}} d\omega
\frac{\omega \, }{\omega^2+\xi^2}{\rm
Im}[f(\omega)(\epsilon(\omega)-1)]\;,\ee and \be J_{\rm
high}^{(p,q)}(\xi)=\frac{2}{\pi \, f({\rm i}
\xi)}\int_{\omega_{\rm max}}^{\infty} d\omega \frac{\omega \,
}{\omega^2+\xi^2}{\rm Im}[f(\omega)(\epsilon(\omega)-1)]\;.\ee
Then by construction we also have: \be \epsilon({\rm i}
\xi)-1=J_{\rm low}^{(p,q)}(\xi)+J_{\rm exp}^{(p,q)}(\xi)+J_{\rm
high}^{(p,q)}(\xi)\;.\ee The quantities $I_{\rm exp}(\xi)$,
$K_{\rm exp}(\xi)$ and $J_{\rm exp}^{(p,q)}(\xi)$ evidently
represent the contribution of the experimental data. On the
contrary the quantities $I_{\rm low}(\xi)$,  $K_{\rm low}(\xi)$
and $J_{\rm low}^{(p,q)}(\xi)$ can be determined only by
extrapolating the data in the low frequency region $0 \le \omega
\le \omega_{\rm min}$, while determination of the quantities
$I_{\rm high}(\xi)$, $K_{\rm high}(\xi)$ and $J_{\rm
high}^{(p,q)}(\xi)$ is only possible after we extrapolate the data
in the high frequency interval $\omega_{\rm max} \le \omega <
\infty$. Ideally, we would like to have $I_{\rm low}(\xi)$,
$I_{\rm high}(\xi)$, $K_{\rm low}(\xi)$, $K_{\rm high}(\xi)$,
$J_{\rm low}^{(p,q)}(\xi)$ and $J_{\rm high}^{(p,q)}(\xi)$ as
small as possible.

To see how things work, we can perform a simple simulation of real
experimental data. We imagine that the electric permittivity of
gold is described by the following six-oscillators approximation
  \cite{Mohid}, which is known to provide a
rather good description of the permittivity of gold for the
frequencies that are relevant to the Casimir effect: \be
\epsilon(\omega)=1-\frac{\omega_p^2}{\omega(\omega+{\rm i}
\gamma)}+\sum_{j=1}^6 \frac{g_j}{\omega_j^2-\omega^2-{\rm i}
\gamma_j \omega}\,.\label{sixosc}\ee Here,  $\omega_p$ is the
plasma frequency and $\gamma$ is the relaxation frequency for
conduction electrons, while the oscillator terms describe core
electrons. The values of the parameters $g_j$, $\omega_j$ and
$\gamma_j$ can be found in the second of Refs. \cite{decca}. For
$\omega_p$ and $\gamma$ we use the reference values for
crystalline  bulk samples, $\omega_p=9$ eV$/\hbar$ and
$\gamma=0.035$ eV$/\hbar$. Of course with such a simple model for
the permittivity of gold, there is no need to use dispersion
relations to obtain the expression of $\epsilon({\rm i}\xi)$, for
this can be simply done by the substitution $\omega \rightarrow
{\rm i} \xi$ in the r.h.s. of Eq. (\ref{sixosc}): \be
\epsilon({\rm i} \xi)=1+\frac{\omega_p^2}{\xi(\xi+
\gamma)}+\sum_{j=1}^6 \frac{g_j}{\omega_j^2+\xi^2+ \gamma_j
\xi}\,.\label{sixoscim}\ee Simulating the real experimental
situation,  let us pretend however that we know that the optical
data of gold are described by Eq. (\ref{sixosc}) only in some
interval $\omega_{\rm min} < \omega < \omega_{\rm max}$, and
assuming that we do not want to make extrapolations of the data
outside the experimental interval, let us see how well the
quantities $I_{\rm exp}(\xi)$, $K_{\rm exp}(\xi)$ and $J_{\rm
exp}^{(p,q)}(\xi)$ defined earlier reconstruct the exact value of
$\epsilon({\rm i} \xi)-1$ given by Eq. (\ref{sixoscim}). In our
simulation we took $\omega_{\rm min }=0.038$ eV$/\hbar$
(representing the minimum frequency value for which data for gold
films were measured in \cite{sveto}) while for $\omega_{\rm max}$
we choose the value $\omega_{\rm max}=30$ eV$/\hbar$. The chosen
value of $\omega_{\rm max}$ is about thirty times the
characteristic frequency $c/(2 a)$ for a separation $a=100$ nm.
The result of our simulation are summarized in Figs. 4 and 5.
\begin{figure}
\includegraphics{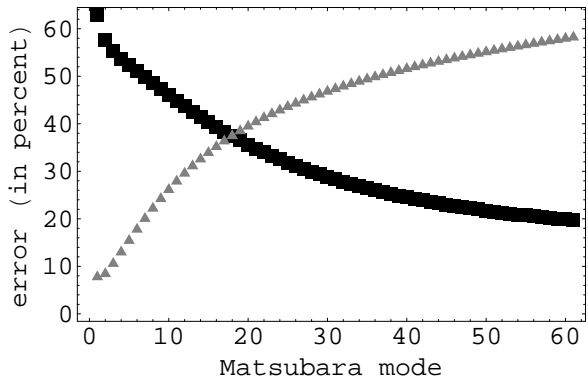}
\caption{\label{fig4} Numerical simulation of the errors (in
percent) in the estimate of $\epsilon({\rm i}\xi_n)-1$ for gold,
resulting from using the quantities $I_{\rm exp}(\xi_n)$ (black
squares) and $K_{\rm exp}(\xi_n)$ (grey triangles) as  estimators,
in the hypothesis that data are available from $\omega_{\rm min
}=0.038$ eV$/\hbar$ to $\omega_{\rm max}=30$ eV$/\hbar$. The
integer on the abscissa labels the Matsubara mode $\xi_n=2 \pi n
k_B T/\hbar$ ($T=300$ K).}
\end{figure}

In Fig. 4, we report the relative per cent errors $\delta_I=100
\,[1-I_{\rm exp}(\xi_n)/(\epsilon({\rm i}\xi_n)-1)]$ (black
squares) and $\delta_K=100 \,[1-K_{\rm exp}(\xi_n)/(\epsilon({\rm
i}\xi_n)-1)]$ (grey triangles) which are made if the quantities
$I_{\rm exp}(\xi_n)$ or  $K_{\rm exp}(\xi_n)$  are used,
respectively, as estimators of $\epsilon({\rm i}\xi_n)-1$. The
integer number on the abscissa labels the Matsubara mode $\xi_n=2
\pi n k_B T/\hbar$ ($T=300$ K). Only the first sixty modes are
displayed, which are sufficient to estimate the Casimir force at
room temperature, for separations larger than 100 nm, with a
precision better than one part in ten thousand. As we see, both
$I_{\rm exp}(\xi_n)$ and $K_{\rm exp}(\xi_n)$ provide a poor
approximation to $\epsilon({\rm i}\xi_n)-1$, with $I_{\rm
exp}(\xi_n)$ performing somehow better at higher imaginary
frequencies, and $K_{\rm exp}(\xi_n)$ doing better at lower
imaginary frequencies. Indeed, $I_{\rm exp}(\xi_n)$ and $K_{\rm
exp}(\xi_n)$ suffer from opposite problems. On one hand the large
error affecting $I_{\rm exp}(\xi_n)$  arises mostly from neglect
of the large low-frequency contribution $I_{\rm low}(\xi_n)$, and
to a much less extent from neglect of the high frequency
contribution $I_{\rm high}(\xi_n)$  (The magnitude of the high
frequency contribution $I_{\rm high}(\xi_n)$ is less than two
percent of $\epsilon({\rm i}\xi_n)-1$ for  all $n \le 60$). The
situation is quite the opposite in the case of $K_{\rm
exp}(\xi_n)$. This difference is of course due to the opposite
limiting behaviors of the imaginary parts of the permittivity
$\epsilon''(\omega)$ in the limits $\omega \rightarrow 0$, and
$\omega \rightarrow \infty$, as compared to those of the imaginary
part of the conductivity $\sigma''(\omega)$. Indeed, for $\omega
\rightarrow 0$, $\epsilon''(\omega)$ diverges like $\omega^{-1}$,
while $\sigma''(\omega)$ approaches zero like $\omega$. This
explains while the low frequency contribution $I_{\rm low}(\xi_n)$
is much larger than $K_{\rm low}(\xi_n)$. On the other hand, in
the limit $\omega \rightarrow \infty$,  $\epsilon''(\omega)$
vanishes like $\omega^{-3}$, while $\sigma''(\omega)$ vanishes
only like $\omega^{-1}$. This implies that large frequencies are
much less of a problem for $I_{\rm exp}(\xi_n)$ than for $K_{\rm
exp}(\xi_n)$. The conclusion to be drawn from these considerations
is that, if either of the two standard forms Eq. (\ref{disp}) or
Eq. (\ref{sigma}) of dispersion relations are used, in order to
obtain a good estimate of $\epsilon({\rm i}\xi_n)-1$, one is
forced to extrapolate somehow the experimental data both to
frequencies less than $\omega_{\rm min}$, and larger than
$\omega_{\rm max}$.

We can now consider our windowed dispersion relation, Eq.
(\ref{generdisp}), with our choice of the window functions $f(z)$
in Eq. (\ref{winfun}). In Fig. 5, we display the relative per cent
error $\delta^{(p,q)}=100 \,[1-J_{\rm
exp}^{(p,q)}(\xi_n)/(\epsilon({\rm i}\xi_n)-1)]$ which is made if
the quantity $J_{\rm exp}^{(p,q)}(\xi_n)$ is used as an estimator
of $\epsilon({\rm i}\xi_n)-1$. We considered two choices of
parameters for our window functions in Eq. (\ref{winfun}), i.e.
$p=1,\,q=2$ (grey triangles) and $p=1,\,q=3$ (black squares). In
both cases, we took for the parameter $w$ the constant value
$w=(1-2\, {\rm i}) \,{\rm eV}/\hbar$ (See Figs. 1, 2 and 3).
\begin{figure}
\includegraphics{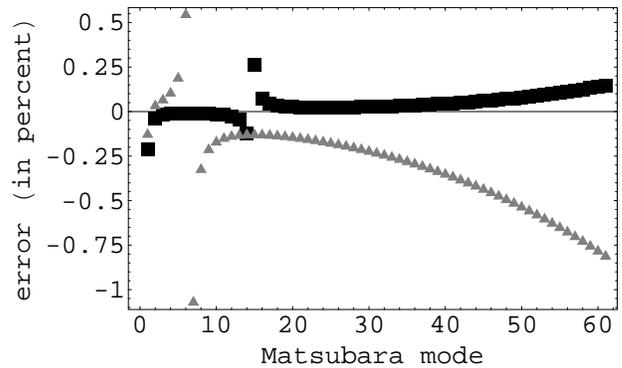}
\caption{\label{fig5} Numerical simulation of the error (in
percent) in the estimate of $\epsilon({\rm i}\xi_n)-1$ for gold,
resulting from using the quantity $J_{\rm exp}^{(p,q)}(\xi_n)$ as
a estimator, in the hypothesis that data are available from
$\omega_{\rm min }=0.038$ eV$/\hbar$ to $\omega_{\rm max}=30$
eV$/\hbar$. The integer on the abscissa labels the Matsubara mode
$\xi_n=2 \pi n k_B T/\hbar$ ($T=300$ K). Grey triangles are for
the window function having $p=1,\,q=2$, black squares for
$p=1,\,q=3$. In both cases $w=(1-2\, {\rm i}) \,{\rm eV}/\hbar$.}
\end{figure}
It is apparent from Fig. 5 that both window functions perform very
well, for all considered Matsubara modes. The error made by using
$J_{\rm exp}^{(1,2)}(\xi_n)$ is less than one percent, in absolute
value, while the error made by using $J_{\rm exp}^{(1,3)}(\xi_n)$
is less than 0.25 percent. The jumps displayed by the relative
errors in Fig. 5 (around $n=6$ for the grey dots, and $n=14$ for
the black ones) correspond to the approximate positions of the
zeroes of the respective window functions $f({\rm i} \xi)$ (see
Fig. 3). Such jumps can be easily avoided, further reducing at the
same time the error, by making a different choice of the free
parameter $w$ for each value of $n$. We did not do this here for
the sake of simplicity. It is clear that in concrete cases one is
free to choose for each value of $n$, different values of all the
parameters $p, q$ and $w$, in such a way that the error is as
small as possible.

\section{Simulation of the Casimir force}

In this Section, we investigate the performance of our window
functions with respect to the determination of the Casimir force.
We consider for simplicity the prototypical case of two identical
plane-parallel homogeneous and isotropic gold plates, placed in
vacuum at a distance $a$. As it is well known, the Casimir force
per unit area is given by the following Lifshitz formula: \be
P(a,T)= \frac{k_B T}{ \pi} \sum_{n \ge 0}{\,'} \int \!\! d{
k_{\perp}} { k_{\perp}} q_n\!\!\!\! \sum_{\alpha={\rm TE,TM}}
\left(\frac{e^{2 a q_n}}{r_{\alpha}^2({\rm i} \xi_n,{ k_{\perp}})}
-1 \right)^{-1},\label{lifs} \ee where the plus sign corresponds
to an attraction between the plates. In this Equation, the prime
over the $n$-sum means that the $n=0$ term has to taken with a
weight one half, $T$ is the temperature, ${ k_{\perp}}$ denotes
the magnitude of the projection of the wave-vector onto the plane
of the plates and $q_n =\sqrt{k_{\perp}^2+\xi_n^2/c^2}$, where
$\xi_n= 2 \pi n\,k_B T /\hbar$ are the Matsubara frequencies. The
quantities $ r_{\alpha}({\rm i} \xi_n,{ k_{\perp}})$ denote the
familiar Fresnel reflection coefficients of the slabs for
$\alpha$-polarization, evaluated at imaginary frequencies $i
\xi_n$. They have the following expressions: \be r_{\rm TE}({\rm
i} \xi_n,{ k_{\perp}})=\frac{q_n-k_n}{q_n+k_n}\;,\ee \be r_{\rm
TM}({\rm  i} \xi_n,{\bf k_{\perp}})=\frac{\epsilon({\rm i}
\xi_n)\, q_n-k_n}{\epsilon({\rm i} \xi_n)\,q_n+k_n}\;,\ee where
$k_n=\sqrt{k_{\perp}^2+\epsilon({\rm i} \xi_n)\xi_n^2/c^2}$.

We have simulated the error made in the estimate of $P(a,T)$ if
the estimate of $\epsilon({\rm i}\xi_n)$ provided by the
window-approximations $J_{\rm exp}^{(p,q)}(\xi_n)$ is used: \be
\epsilon({\rm i}\xi_n) \simeq 1+ J_{\rm exp}^{(p,q)}(\xi_n)\;,\ee
again assuming the simple six-oscillator model of Eq.
(\ref{sixosc}) for $\epsilon(\omega)$. The results are summarized
in Fig 6, where we plot the relative error $\delta_P^{(p,q)}$ in
percent, as a function of the separation $a$ (in microns). The
window functions that have been used are the same as in Fig. 5.
\begin{figure}
\includegraphics{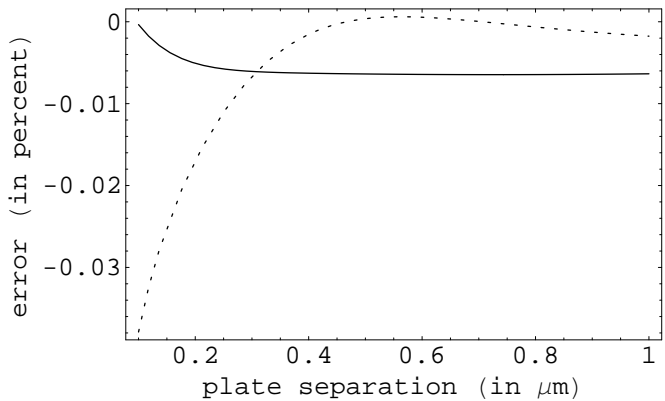}
\caption{\label{fig6} Simulation of the   error (in percent),
versus plate separation (in $\mu$m) in the estimate of the Casimir
force per unit area, between two plane-parallel gold plates in
vacuum at a temperature $T=300 K$, resulting from using $J_{\rm
exp}^{(p,q)}(\xi_n)$ as an estimator of  $\epsilon({\rm
i}\xi_n)-1$. The window functions are the same as in Fig. 5: the
dashed line is for $p=1, q=2$ and the solid line for $p=1, q=3$.
All values of the other parameters are same as in Fig. 5.}
\end{figure}
We see from the figure that already with this simple and
not-optimized choice of window functions, the error is much less
than one part in a thousand  in the entire range of separations
considered, from 100 nm to one micron.

\section{Conclusions and discussion}

In recent years, a lot of efforts have been made to measure
accurately the Casimir force. At the moment of this writing, the
most precise experiments using gold-coated micromechanical
oscillators claim a precision better than one percent
\cite{decca}. It is therefore important to see if an analogous
level of precision in the prediction of the Casimir force can be
obtained at the theoretical level. A precise determination of the
theoretical error is indeed as important as reducing the
experimental error, in order to address controversial questions
that have emerged in the recent literature on dispersion forces,
regarding the influence of free charges on the thermal correction
to the Casimir force \cite{Mohid}.

Addressing the theoretical error in the magnitude of the  Casimir
force  is indeed difficult, because many physical effect must be
accounted for. However, it has recently been pointed out
\cite{sveto} that perhaps the largest theoretical uncertainty
results from incomplete knowledge of the optical data for the
surfaces involved in the experiments. On one hand, the large
variability depending on the preparation procedure, of the optical
properties of gold coatings, routinely used in Casimir
experiments, makes it necessary to accurately characterize the
coatings actually used in any experiment. On the other hand, even
when this characterization is done, another problem arises,
because for  evaluating the Casimir force one needs to determine
the electric permittivity $\epsilon({\rm i} \xi)$ of the coatings
for certain imaginary frequencies ${\rm i} \xi$. This quantity is
not directly accessible to any optical measurement, and the only
way to determine it is via exploiting dispersion relations, that
permit to express $\epsilon({\rm i} \xi)$ in terms of the
measurable values of the permittivity $\epsilon(\omega)$ for real
frequencies $\omega$. When doing this, one is faced with the
difficulty that optical data are necessarily known only in a
finite interval of frequencies $\omega_{\rm min} < \omega <
\omega_{\rm max}$. This practical limitation constitutes a severe
problem in the experimentally relevant case of good conductors,
because of their large conductivity at low frequencies.  With the
standard forms of dispersion relations Eq. (\ref{disp}) and Eq.
(\ref{sigma}), one finds that for practical values of $\omega_{\rm
min}$ and $\omega_{\rm max}$, low frequencies less than
$\omega_{\rm min}$ and/or large frequencies larger than
$\omega_{\rm max}$ give a very large contribution to
$\epsilon({\rm i} \xi)$. In order to estimate $\epsilon({\rm i}
\xi)$ accurately, one is then forced to extrapolate available
optical data outside the experimental region, on the basis of some
theoretical model for $\epsilon(\omega)$. Of course, this
introduces a further element of uncertainty in the obtained values
of $\epsilon({\rm i} \xi)$, and the resulting theoretical error is
difficult to estimate quantitatively.

In this paper we have shown that this problem can be resolved by
suitably modifying the standard dispersion relation used to
compute $\epsilon({\rm i} \xi)$, in terms of appropriate analytic
window  functions $f(z)$ that suppress the contributions both of
low and large frequencies. In this way, it becomes possible to
accurately estimate $\epsilon({\rm i} \xi)$ solely on the basis of
the available optical data, rendering unnecessary any
uncontrollable extrapolation of data. We have checked numerically
the performance of simple choices of window functions, by making a
numerical simulation based on an analytic fit of the optical
properties of gold, that has been used in recent experiments on
the Casimir effect \cite{Mohid}. We found that already very simple
forms of the window functions permit  to estimate the Casimir
pressure with an accuracy better than one part in a thousand, on
the basis of reasonable intervals of frequencies for the optical
data. It would be interesting to apply these methods to the
accurate optical data for thin gold films quoted in Ref.
\cite{sveto}.

Before closing the paper, we should note  that the relevance of
the sample-to-sample dependence of the optical data observed in
\cite{sveto} for the theory of the Casimir effect has been
questioned by  the  authors of Ref. \cite{Mohid}, who observed
that this dependence  mostly originates from relaxation processes
of free conduction electrons at infrared and optical frequencies,
due for example to different grain sizes in thin films. The main
consequence of these sample-dependent features  is the large
variability of the Drude parameters, extracted from  fits of the
low-frequency optical data of the films, which constitutes the
basic source of variation of the computed Casimir force  reported
in Ref. \cite{sveto}. According to the authors of Ref.
\cite{Mohid}, relaxation properties of conduction electrons in
thin films, described  by the fitted values of the Drude
parameters, are not relevant for the Casimir effect. Indeed,
according to these authors the quantity $\epsilon(\omega)$ to be
used in Lifshitz formula should not be understood as the actual
electric permittivity of the plate, as derived from optical
measurements on the sample, but it should be rather regarded as a
phenomenological quantity connected to but not identical to the
optical electric permittivity of the film. The ansatz offered by
them for $\epsilon(\omega)$ is dubbed as generalized plasma model,
and following Ref. \cite{Mohid} we denote it as $\epsilon_{\rm
gp}(\omega)$. This quantity is a semianalytical mathematical
construct, defined by the formula: \be \epsilon_{\rm
gp}(\omega)=\epsilon_c(\omega)-\omega_p^2/\omega^2\;,\label{genpla}\ee
where $\epsilon_c(\omega)$ represents the contribution of core
electrons, while the term proportional to the square of the plasma
frequency $\omega_p$ describes conduction electrons. The most
striking qualitative feature of this expression is the neglect of
ohmic dissipation in the contribution from conduction electrons,
but this is not all. Indeed, the ansatz prescribes that only the
core-electron contribution $\epsilon_c(\omega)$ should be
extracted from optical data of the film. On the contrary, and more
importantly, according to Ref. \cite{Mohid} the value of the
plasma frequency $\omega_p$ to be used  in Eq. (\ref{genpla})
should be the one pertaining to a perfect crystal of the ${\it
bulk}$ material, and not the one obtained by a Drude-model fit of
the low-frequency optical data of the film actually used in the
experiment. The justification provided for this choice of the
plasma frequency by the authors of Ref. \cite{Mohid} is that the
contribution of conduction electrons to the Casimir force should
depend only on properties determined by the structure of the
crystal cell, which are independent   of the sample-to-sample
variability determined by the peculiar grain structure of the
film, reported in Ref. \cite{sveto}. It should be noted that for
gold, the value of the plasma frequency advocated in \cite{Mohid},
$\omega_p=9$ eV/$\hbar$, is much higher than the fit values quoted
in Ref. \cite{sveto}, which range from 6.8 to 8.4 eV/$\hbar$. As a
result, the approach advocated in Ref. \cite{Mohid}  leads to
larger magnitudes of the Casimir force, as compared to the values
derived in Ref. \cite{sveto}, with differences ranging, depending
on the sample, from 5 $\%$ to 14 $\%$ at 100 nm. There is no room
here to further discuss  the merits and faults of these
approaches, and we refer the reader to \cite{Mohid} for a thorough
analysis. It is fair to note though that a series of recent
experiments by one experimental group \cite{decca} appears to
favor the generalized plasma approach, and to rule out the more
conventional approach based on actual optical data followed in
Refs. \cite{lamor2,sveto}.

The future will tell what is the correct description. In the
meanwhile, we remark that whatever approach is followed, the
methods proposed in this paper may prove useful to obtain more
reliable estimates of the Casimir force for future experiments.

\noindent {\it Acknowledgements} The author thanks the ESF
Research Network CASIMIR for financial support.

\end{document}